\documentclass[structabstract]{aa}  
\usepackage{graphicx}
\usepackage{epsfig}
\usepackage{txfonts}
\begin{document}
\authorrunning{T. Johannsen}
\titlerunning{Constraints on the Size of Extra Dimensions from XTE J1118+480}
   \title{Constraints on the Size of Extra Dimensions from the Orbital Evolution of the Black-Hole X-Ray Binary XTE J1118+480}

   \author{T. Johannsen
          }

   \institute{Physics Department, The University of Arizona, 1118 E. 4th Street, Tucson, AZ 85721, USA\\
              \email{timj@physics.arizona.edu}
            }

   \date{Received June 30, 2009}

 
  \abstract
   {To constrain Randall-Sundrum type braneworld gravity models and the expected rapid evaporation of astrophysical black holes due to the emission of gravitational modes in the extra dimension.}
   {It is argued that the black-hole X-ray binary XTE J1118+480 is suitable for a constraint on the asymptotic curvature radius of the extra dimension in such braneworld models. An upper limit on the rate of change of the orbital period of XTE J1118+480 is obtained.}
   {The expected black-hole evaporation in the extra dimension leads to a potentially observable rate of change of the orbital period in XTE J1118+480. The time-change of the orbital period is calculated from previous orbital period measurements from the literature. The lack of observed orbital period evolution is used to constrain the asymptotic curvature radius of the extra dimension.}
   {The asymptotic AdS radius of curvature is constrained to a value comparable to other limits from astrophysical sources. A unique property of XTE J1118+480 is that the expected rate of change of the orbital period due to magnetic braking alone is so large that only one additional measurement of the orbital period would lead to the first detection of orbital evolution in a black-hole binary and impose the tightest constraint to date on the size of one extra dimension of the order of $35{\rm \mu m}$.}
   {}


   \maketitle
%

\section{Introduction}

The so-called hierarchy problem is one of the key open questions in the search for a grand theory beyond general relativity and the standard model. The fundamental scale of Einsteinian gravity, the Planck scale, differs from the electroweak scale by 16 orders of magnitude, thus making general relativity and the standard model hard to reconcile in a grand unification sceme. Braneworld Gravity, as a candidate for a unified theory, requires the existence of more than three spatial dimensions (see Maartens 2004 for a review), but the size of a potential extra dimension has already been constrained to the sub-mm range by precision tests of Newton's inverse square law (see, e.g., Adelberger, Heckel \& Nelson 2003 for a review).

 One possible solution (Arkani-Hamed, Dimopoulos \& Dvali 1998) is to demand that extra dimensions be compactified at a scale smaller than those probed by experiment. All standard-model particles are bound to the known four-dimensional world (``the brane''). Only gravity can access the extra dimensions, which accounts for its apparent weakness, because it is spread out through higher-dimensional space (``the bulk''). This approach reconciles the gravitational and the electroweak scales if the extra dimensions are large enough, but it is beyond the scope of astrophysical tests.

A second scenario (Randall \& Sundrum 1999) embeds the brane in a five-dimensional anti-de Sitter space, which allows for the extra dimension to be infinite. The bulk is filled with a negative cosmological constant such that the extra dimension only effects the brane on a length scale which is small enough and which is set by the asymptotic curvature radius $L$ of the bulk. This model has striking consequences for astrophysical black holes.

Perturbative solution of the classical bulk equations (Tanaka 2003) indicated that no stable black holes can exist on the brane. A treatment of higher-dimensional black holes via the AdS/CFT correspondence showed that black holes are indeed unstable and lose energy in the extra dimension through the emission of CFT modes (Emparan, Garc\'{i}a-Bellido \& Kaloper 2003; see, however, Fitzpatrick, Randall \& Wiseman 2006). The lifetimes of astrophysical black holes are dramatically reduced and can be as short as a Megayear if the asymptotic curvature $L$ of the extra dimension turns out to be in the sub-mm range.

For a binary system consisting of a black hole and a companion star, this effect should be observable and lead to a measurable change of the orbital period (Johannsen, Psaltis \& McClintock 2009). Several black-hole binaries were identified as candidates, and the system SXT A0620-00 yielded a constraint on the asymptotic curvature radius of $L<161~{\rm \mu m}$ at 3$\sigma$ (Johannsen et al. 2009). Similar limits have been obtained from the age of the black hole XTE J1118+480 ($L<80~{\rm\mu m}$; Psaltis 2007) and from tabletop experiments of Newton's inverse square law (Adelberger et al. 2007; Geraci et al. 2008). The current 3$\sigma$-upper limit on the AdS radius $L$ is of the order of $44~{\rm\mu m}$ (Kapner et al. 2007).

In this paper, I extend the analysis of Johannsen et al. (2009) to the black-hole binary XTE J1118+480 and compute an additional constraint on the asymptotic curvature radius $L$. In \S 2, I briefly review the results of Johannsen et al. (2009) for the evolution of the orbital period of a black-hole binary system with non-conservative mass transfer and black-hole evaporation. In \S 3, I apply this result to the black-hole binary J1118+480 and I obtain a constraint on the asymptotic curvature radius of $L<0.97~{\rm mm}$ in \S 4.


\section{Black-Hole Binaries in Braneworld Gravity}

For a binary system with a black hole of mass $m_1$ and a companion star of mass $m_2$ on a circular orbit, the orbital angular momentum, $J\equiv\mu\sqrt{Gma}$, is subject to change because of magnetic braking (e.g., Webbink, Rappaport \& Savonije 1983) and the evolution of the secondary star (e.g., Verbunt 1993). Here, $m\equiv m_{1}+m_{2}$, $\mu\equiv m_{1}m_{2}/m$, and $a$ is the 
semi-major axis. A third effect is the emission of CFT modes in the extra dimension (Johannsen et al. 2009), where the black-hole mass loss is given by (Emparan et al. 2003)\begin{equation}
\dot{M}=2.8\times10^{-3}\left(\frac{M_\odot}{m_{1}}\right)^{2}\left(\frac{L}{1~{\rm mm}}\right)^{2}M_\odot\, {\rm yr^{-1}}.\label{Mdot}\end{equation}
Here, $L$ is the asymptotic AdS radius of curvature.

This leads to a change of the orbital period of (Johannsen et al. 2009)\[
\frac{\dot{P}}{P}=Q_{1}\left(\frac{M_\odot}{m_{1}}\right)^{3}\left(\frac{L}{1~{\rm mm}}\right)^{2}\]
\[
+Q_{2}\frac{(m_{1}+m_{2})^{2}}{m_{1}}
\left[\frac{0.49q^{-2/3}}{0.6q^{-2/3}+\ln(1+q^{-1/3})}\right]^{\gamma}
\left[\frac{\sqrt{G(m_{1}+m_{2})}}{2\pi}P\right]^{\frac{2}{3}(\gamma-5)}\]
\begin{equation}
+Q_{3}\left(c_{1}+2c_{2}y+3c_{3}y^{2}\right)e^{a_{0}+a_{2}y^{2}+a_{3}y^{3}}\left(\frac{M_{{\rm c}}}{M_\odot}\right)^{a_{1}-1},\label{final}\end{equation}
where $q\equiv m_1/m_2$, and $c_1$, $c_2$, $c_3$, as well as $a_0$, $a_1$, $a_2$, and $a_3$ are constants depending on the core composition of the companion star. The parameter $\gamma$ governs the strength of magnetic braking (c.f. eq. (36) in Rappaport, Verbunt \& Joss 1983). The additional quantities are defined by (Johannsen et al. 2009)

\[
Q_{0}\equiv\frac{1}{2}\frac{1-\beta}{1+q}\frac{1+\frac{1}{2}\frac{q}{1+q}+\frac{1}{3}\mathcal{A}}{D}+\frac{1}{2}\frac{q}{1+q}\]
\[+\frac{3}{2}\frac{\left(\frac{2}{3}\frac{\beta+q}{q}\mathcal{A}-\xi_{{\rm ad}}\right)\left(1+\frac{1}{2}\frac{q}{1+q}+\frac{1}{3}\mathcal{A}\right)}{D}-\mathcal{A},\]
\[Q_{1}\equiv2.8\times10^{-3}Q_{0}\, {\rm yr^{-1}},\]
\[Q_{2}\equiv-\frac{3.8\times10^{-30}GR_\odot^{4-\gamma}}{D}\left(\frac{1}{2}\frac{1-\beta}{1+q}+\frac{\beta+q}{q}\mathcal{A}-\frac{3}{2}\xi_{{\rm ad}}\right),\]
\[Q_{3}\equiv1.37\times10^{-11}\times4^{a_{1}}
\left[\frac{1}{4D}\frac{1-\beta}{1+q}+\frac{1}{2D}\left(\frac{\beta+q}{q}\mathcal{A}-\frac{3}{2}\xi_{{\rm ad}}\right)-\frac{3}{2}\right], \]
\[\mathcal{A}\equiv1-\frac{0.6+0.5q^{1/3}(1+q^{-1/3})^{-1}}{0.6+q^{2/3}\ln\left(1+q^{-1/3}\right)},\label{A} \]
\begin{equation}D\equiv j_{{\rm w}}(1-\beta)\frac{1+q}{q}-1+\frac{\beta}{q}+\frac{1}{2}\frac{1-\beta}{1+q}-\frac{1}{2}\left(\xi_{{\rm ad}}-\frac{2}{3}\frac{\beta+q}{q}\mathcal{A}\right).\label{D} \end{equation}

\noindent Here, $P$ is the orbital period, $\beta$ is the fraction of matter that is accreted by the black hole, $\xi_{{\rm ad}}$ is the adiabatic index of the companion star, and $j_{\rm w}$ is the specific angular momentum in units of $2\pi a^2/P$ that is lost through the stellar wind, which carries away angular momentum $J$ at a rate (Will \& Zaglauer 1989)
\begin{equation}
\frac{\dot{J}_{\rm w}}{J}=j_{\rm w}(1-\beta)\frac{1+q}{q}\frac{\dot{m}_2}{m_2}.
\end{equation}
\noindent In this expression, $\dot{J}_{\rm w}$ is the angular momentum loss due to stellar wind, $m_1$ and $m_2$ are the masses of the black hole and the companion star, respectively, $\dot{m}_1$ and $\dot{m}_2$ their time derivatives, and $q=m_1/m_2$ is the mass ratio.

Observations of the system A0620-00 have been used previously in conjunction with the above theoretical prediction to constrain the asymptotic curvature radius to a value of $L\leq161~{\rm\mu m}$ (Johannsen et al. 2009). In the following, I  apply the formalism above to the system XTE J1118+480.

\section{The Orbital Evolution of XTE J1118+480} 

In order to obtain a constraint on the asymptotic curvature radius $L$ of the extra dimension, it is essential to select a black-hole binary with an unevolved companion star. In that case, the evolution term in equation (\ref{final}) can be neglected, and the binary can be used to constrain the AdS radius as long as the magnetic braking term is negligible compared to the evaporation term. In the following I argue that this approach can be applied to the black-hole binary J1118+480, and I use previous measurements of its orbital period to place a bound on the rate of change of its orbital period.

The system J1118+480 has been monitored for more than a decade (Remillard \& McClintock 2006). The companion star of the black hole resembles a late-type main-sequence star of spectral type K7 V -- M0 V (Wagner et al. 2001). In addition, the mean density is only $\sim$50\% higher than for a usual main-sequence M0 star (McClintock et al. 2001), and the mass is only $\sim$50\% lower than that of such a star (see Charles \& Coe 2006).

It is important to note that this is not simply a normal star. It has emerged out of an exeptional evolutionary history (see de Kool et al. 1986 for an example), and it does not evolve on a nuclear timescale. For my analysis, however, it is sufficient that the secondary only behaves like a main-sequence star. Then the evolution term in equation (\ref{final}) is negligible.

Considering only the evaporation term and the magnetic braking term in equation (\ref{final}), I plot in Figure~1 the rate of change of the orbital period versus the asymptotic curvature radius $L$ in the extra dimension for the binary systems J1118+480 and A0620-00. The parameters used in this plot are $\xi_{{\rm ad}}=0.8$, $\beta=0$, $j_{\rm w}=0$, and $\gamma=0$. For values of the asymptotic curvature radius greater than $L\simeq35~{\rm\mu m}$ the evaporation term dominates the evolution of the orbital period in the case of J1118+480. Below that value, the magnetic braking is predominant. For the binary A0620-00 the transition occurs at $L\simeq20~{\rm\mu m}$ (Johannsen et al. 2009). Consequently, these sources are similar in constraining the asymptotic curvature radius of the extra dimension.
 
\begin{figure}
\centering
\includegraphics[width=1.0\linewidth]{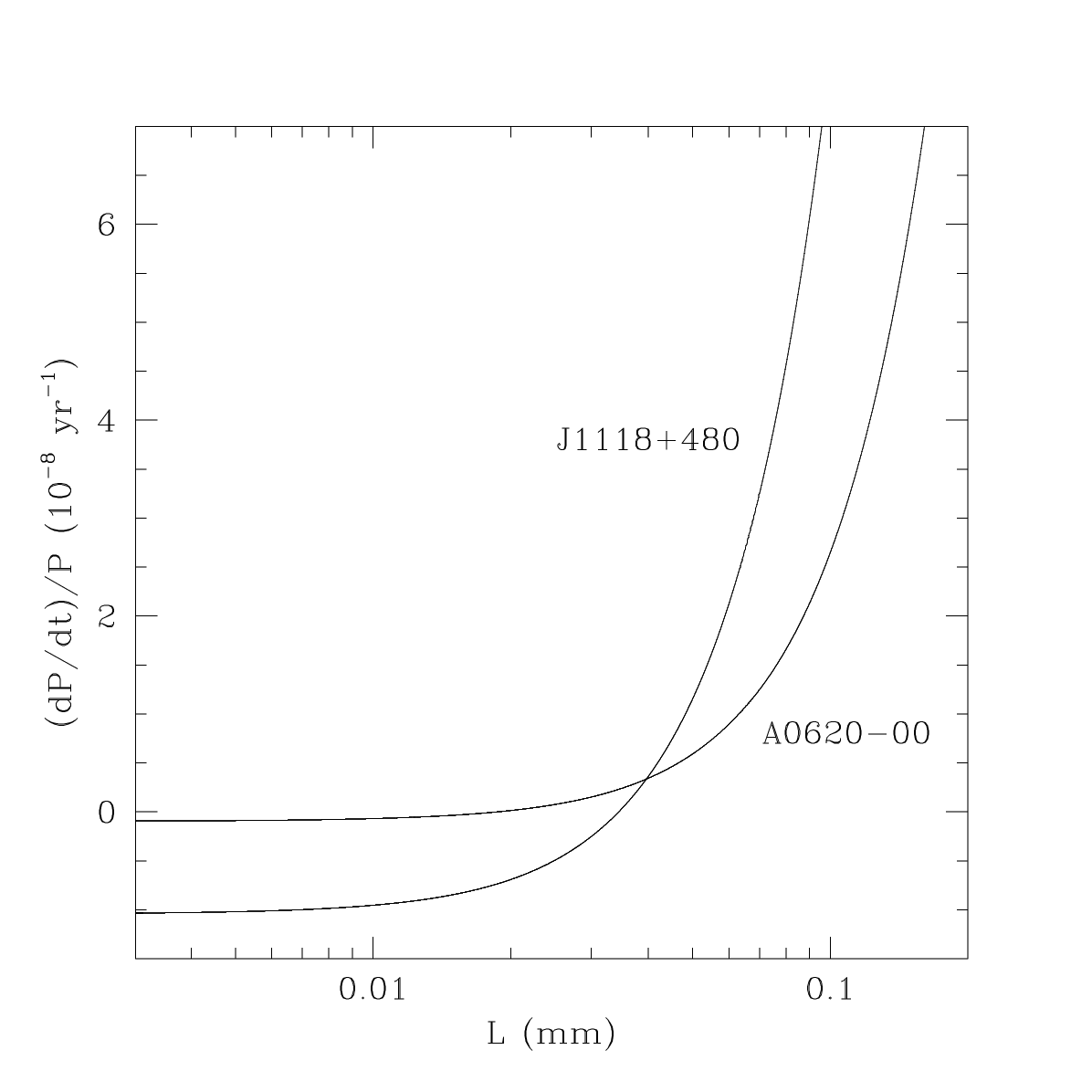}
\caption{The rate of change of the orbital period $P$ of the binary systems J1118+480 and A0620-00 versus the asymptotic curvature radius $L$ in the extra dimension. The parameters are $\xi_{{\rm ad}}=0.8$, $\beta=0$, $j_{\rm w}=0$, and $\gamma=0$. The transition from predominant magnetic braking (constant rate) to predominant black-hole evaporation (rapidly increasing rate) occurs at $L\simeq20~\mu{\rm m}$ (A0620-00) and $L\simeq35~\mu{\rm m}$ (J1118+480), respectively.\label{fig2}}
\end{figure}

A measurement of a positive rate of change of the orbital period automatically constrains the asymptotic curvature radius to an interval, as can be seen by simply evaluating the magnetic braking term. For a binary system with a high mass ratio, i.e. $m_1/m_2>5.5$, the effect of magnetic braking can only decrease the orbital period (Johannsen et al. 2009). Thus such a measurement actually determines the AdS radius $L$ and would prove the existence of an extra dimension.

In Figure~2, I plot the rate of change of the orbital period of J1118+480 as a function of the parameters  $j_{\rm w}$, $\beta$, and $\gamma$. On varying one of them, the others are held constant at the respective values $\beta=0$, $j_{\rm w}=0$, and $\gamma=0$. For all plots I choose $\xi_{{\rm ad}}=0.8$ and $L=44~\mu{\rm m}$, which is the value of the current experimental limit (Kapner et al. 2007). I am interested in the smallest rate of orbital period evolution, so in the following I set the parameters to the respective values $j_{\rm w}=0$ (no angular momentum loss due to stellar wind), $\beta=0$ (no accretion onto the black hole), and $\gamma=0$.

\begin{figure*}
\begin{center}
\psfig{figure=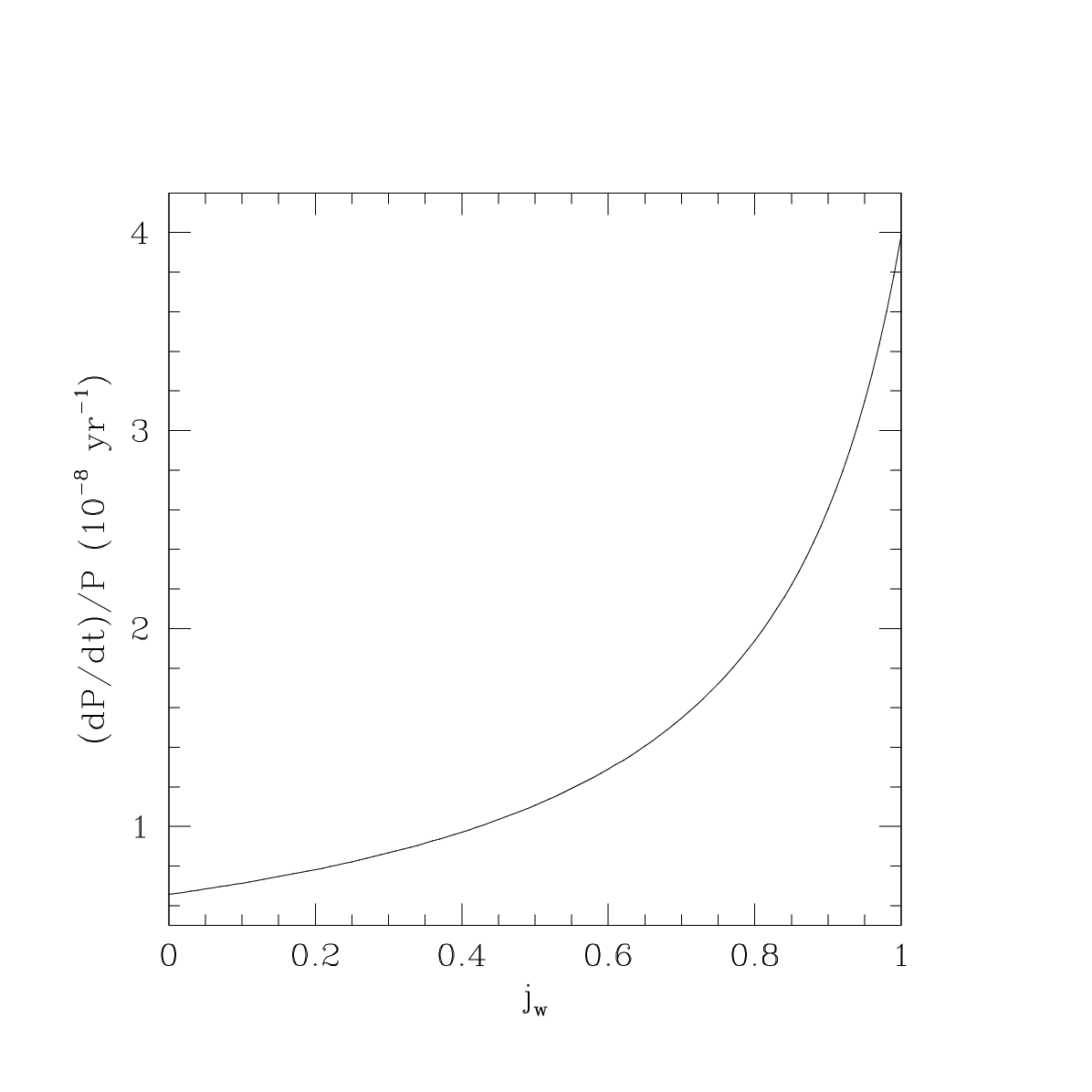,width=5.7cm}
\psfig{figure=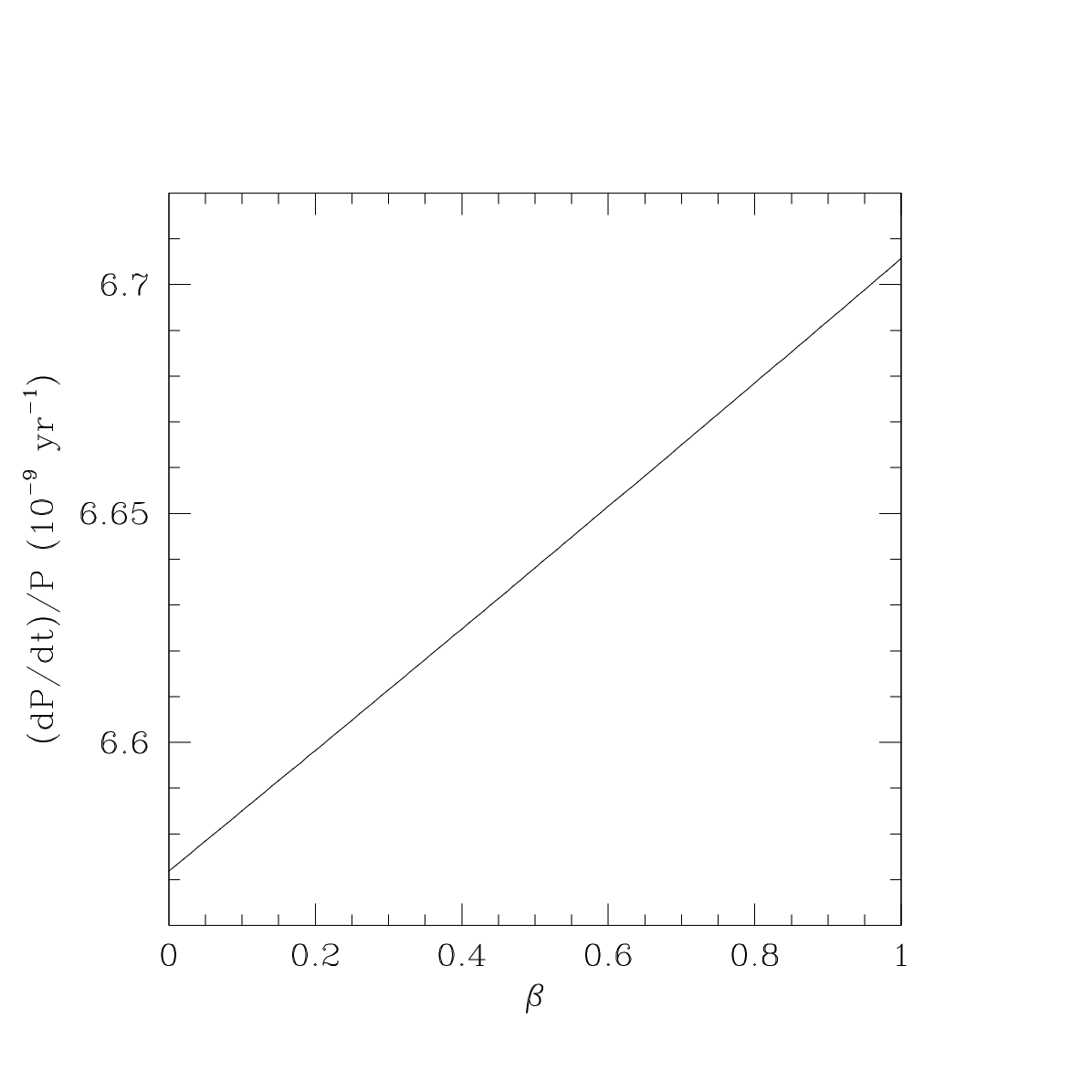,width=5.7cm}
\psfig{figure=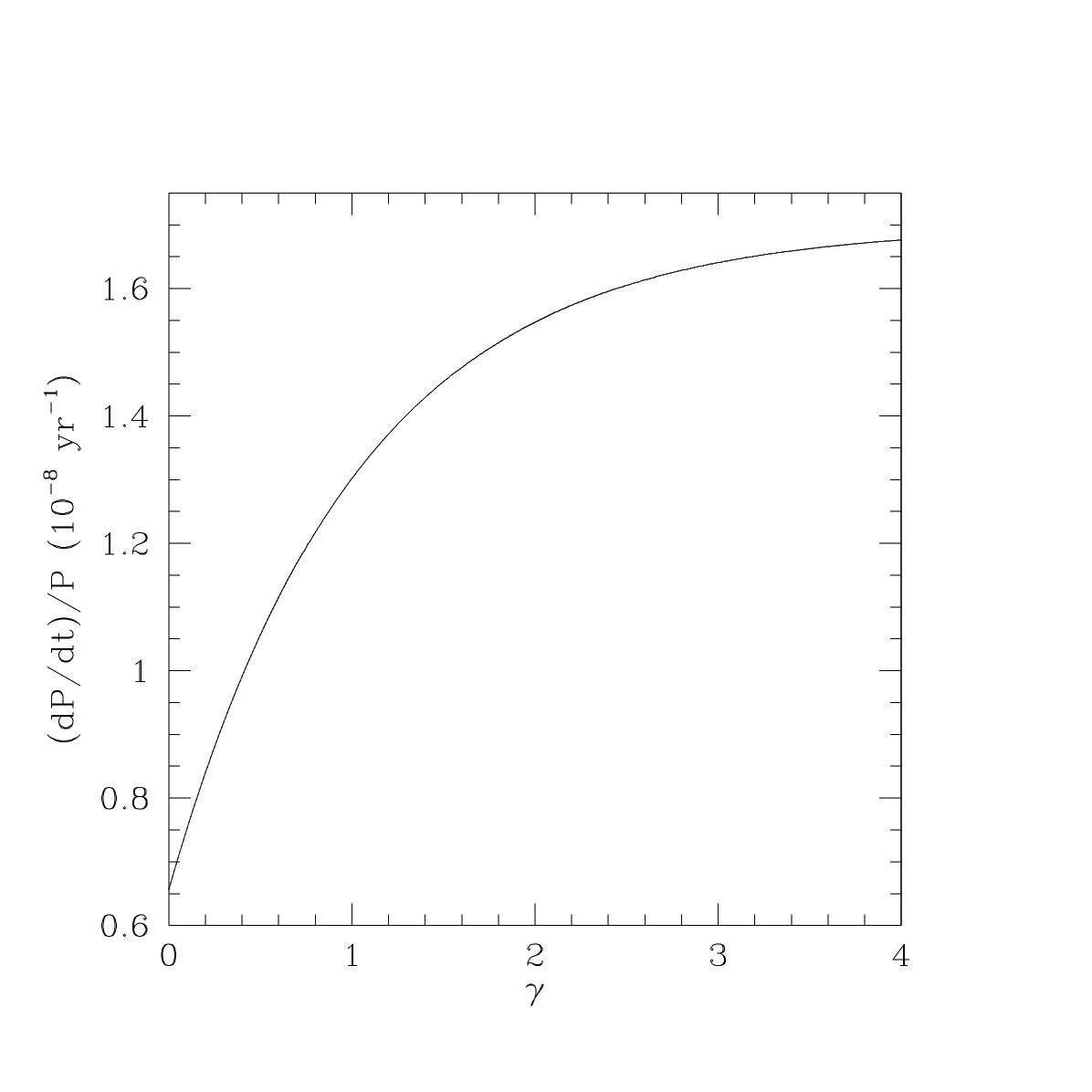,width=5.7cm}
\caption{The rate of change of the orbital period $P$ (in years) of the binary system J1118+480 versus the specific angular momentum removed by the stellar wind $j_{\rm w}$, the accretion parameter $\beta$, and the magnetic braking parameter $\gamma$, for $L=44~\mu{\rm m}$ and $\xi_{\rm ad}=0.8$. On varying one parameter, the others are held constant at the respective values $j_{\rm w}=0$, $\beta=0$, and $\gamma=0$.}
\end{center}
\end{figure*}

The orbital period $P$ of J1118+480 has been measured several times over the past decade. Based on those measurements, I calculate the rate of change of the orbital period. Table 1 shows the orbital period measurements for J1118+480. The values are for the most part consistent with each other, indicating an at most small change of the period over the last years.

\begin{table*}[ht]
\begin{center}
\footnotesize
\begin{tabular}{lcccl}
\multicolumn{5}{c}{Table 1: The Observed Orbital Periods and Times of 
Measurement for J1118+480}\\
\hline \hline
$P\ (d)$ & $T_0$ (inf. conj.) in HJD & Orbital Cycle $n$ & Reference$^{\rm a}$\\
\hline
0.169930$\pm$0.000004 & 2,451,868.8916$\pm$0.0004 & 0 & 1\\
0.17013$\pm$0.00010 & 2,451,880.1485$\pm$0.0010$^{\rm b}$ & 66 & 2\\
0.1699339$\pm$0.0000002 & 2,451,880.1086$\pm$0.0004& 66 & 3\\
0.169937$\pm$0.000001 & 2,452,022.5122$\pm$0.0004 & 904 & 4\\
0.16995$\pm$0.00012 & 2,453,049.93346$\pm$0.00007 & 6950 & 5\\
\hline
\end{tabular}
\item[] $^{\rm a}$ (1) Wagner et al. 2001; (2) McClintock et al. 2001; (3) Torres et al. 2004; (4) Zurita et al. 2002; (5) Gonz\'{a}lez Hern\'{a}ndez et al. 2008
\item[] $^{\rm b}$ $T_0$ time of maximum velocity
\end{center}
\end{table*}

Assuming a constant rate of change of the orbital period, the time of the $n$th
orbital cycle is given by (e.g., Kelley et al. 1983)

\begin{equation}
t_{\rm n} = t_{\rm 0} + Pn+\frac{1}{2}P\dot{P}n^2,\label{Pdotexp}
\end{equation}

\noindent where $P$ is the orbital period at time $t_{\rm 0}$, $\dot{P}$ is its derivative, and $n$ is the orbital cycle number. In order to calculate the rate of change of the orbital period, I fitted the times $T_0$ in Table 1 using the IDL routine {\it curvefit}. The value from reference (2) was omitted, and instead the more precise reference (3) was used. The fit yields $\dot{P}=(0.7\pm3.1)\times10^{-9}~{\rm s/s}$ which is still consistent with zero quoting 3$\sigma$-errors. The fit and residuals are shown in Figure~3.

\begin{figure}
\begin{center}
\includegraphics[width=1.0\linewidth]{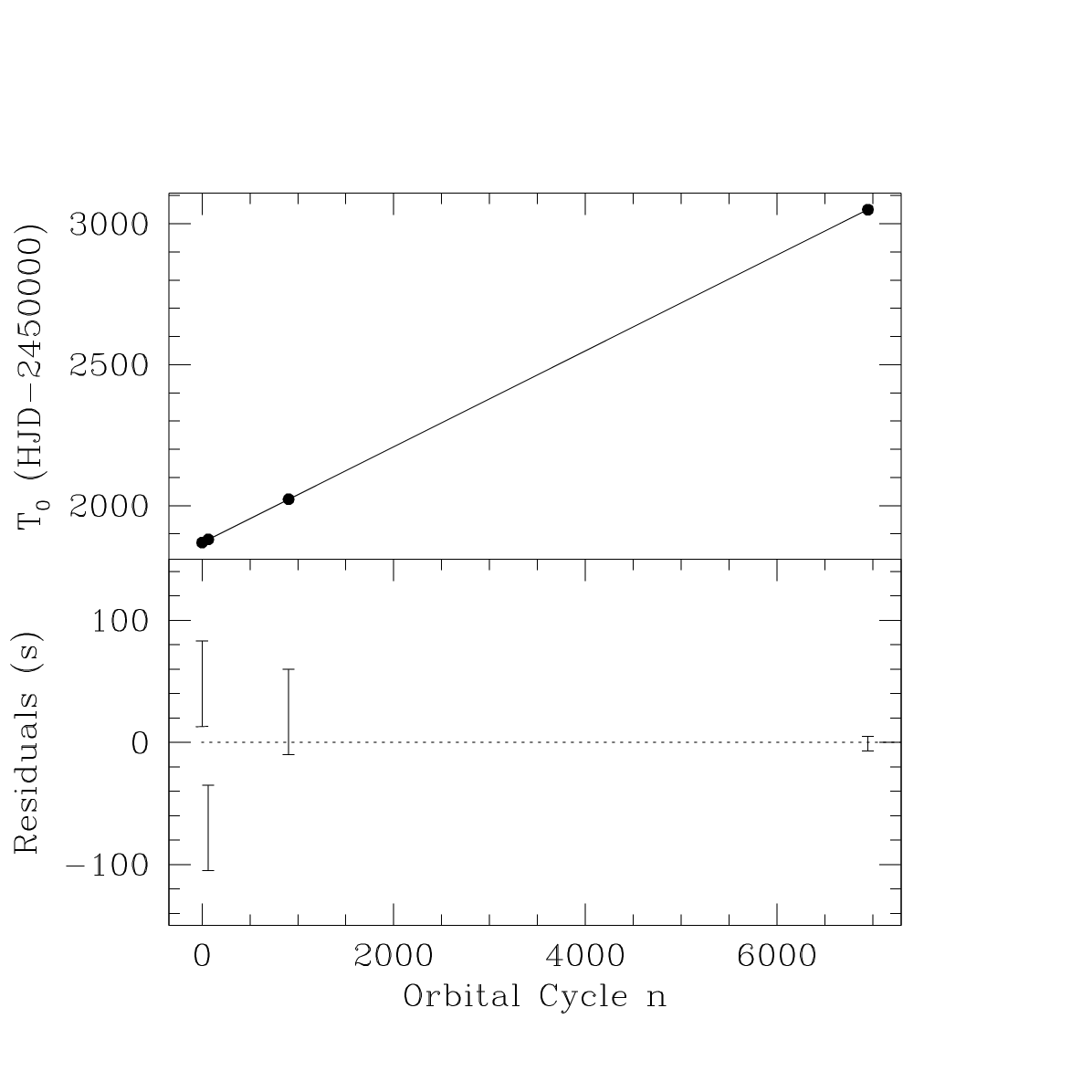}
\caption{The times $T_0$ of the inferior conjunction and residuals versus the orbital cycle number $n$ for the black-hole binary J1118+480.\label{fig4}}
\end{center}
\end{figure}

\begin{figure}
\begin{center}
\includegraphics[width=1.0\linewidth]{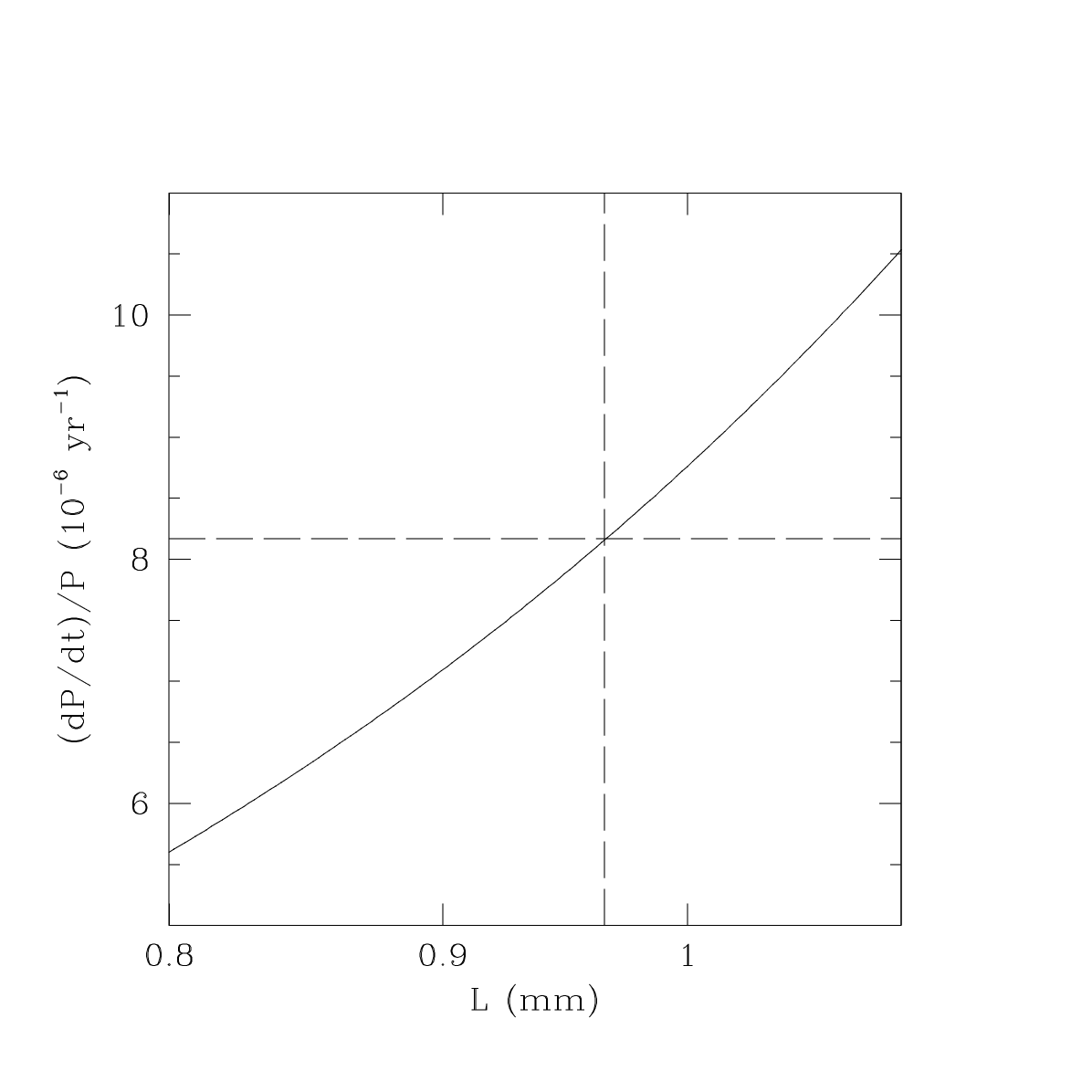}
\caption{The minimum rate of change of the orbital period $P$ of the binary J1118+480 versus the asymptotic curvature radius $L$ in the extra dimension. The intersection point of the observed 3$\sigma$-upper limit on $\dot{P}/P$ (horizontal line) with the minimum rate of change of the orbital period marks the constraint on the AdS curvature radius of $L<0.97~{\rm mm}$ (vertical line).\label{fig5}}
\end{center}
\end{figure}

\section{Results}

Using the values of the orbital period derivative calculated in \S 3, I compute a limit on the asymptotic curvature radius of the extra dimension. For the mass of the black hole I use the best fit value of $m_1=6.8~M_{\odot}$ (c.f. Charles \& Coe 2006).

The 3$\sigma$-upper limit on the observed rate of change of the orbital period and the smallest theoretical expectation thereof determine an upper bound on the AdS curvature radius $L$. In Figure~4, I plot the rate of change of the orbital period of J1118+480 for the set of parameters that minimizes it (c.f. \S 3). The intersection point of this curve with the 3$\sigma$-upper limit on the orbital period evolution yields a constraint on the AdS curvature radius of $L<0.97~{\rm mm}$.

This value is comparable to the constraint obtained from A0620-00 (Johannsen et al. 2009). A refinement of that upper bound would be easy to obtain and simply requires an additional measurement of the orbital period. The fact that the period change due to magnetic braking is so large for J1118+480 in the case where the black-hole evaporation is negligible ($\dot{P}/P\sim -10^{-8}~{\rm yr^{-1}}$; c.f. Figure~1) has two important consequences. First, even one additional measurement of the orbital period in 2009 ($n=17,000$) with an error in the ephemeris $T_0$ of the order of $0.001{\rm d}$ would prove the period derivative to be negative at $3\sigma$, hence measuring the first orbital period evolution of a black-hole X-ray binary. Second, such a measurement would show that black-hole evaporation is negligible against magnetic braking in this binary, which would in turn impose a constraint of the order of $35~{\rm \mu m}$ on the size of one extra dimension, the tightest constraint to date.

The constraint on the asymptotic curvature radius also depends on the mass of the primary, but since the black-hole mass has been measured quite precisely ($m_1=6.8\pm0.4~M_{\odot}$; from Charles \& Coe 2006), its effect is small.

\section{Conclusions}

   \begin{enumerate}
      \item The black-hole binary XTE J1118+480 is well-suited for a constraint on the asymptotic curvature radius of the extra dimension in Randall-Sundrum type braneworld gravity models.
      \item An upper limit on the rate of change of the orbital period of XTE J1118+480 is calculated based on previous measurements of the orbital period from the literature. The lack of observed orbital evolution of this binary imposes a constraint on the asymptotic curvature radius of $L<0.97~{\rm mm}$.
      \item This constraint can be significantly improved by only one additional measurement of the orbital period of XTE J1118+480. This would be the first detection of orbital evolution in a black-hole binary. The expected predominance of magnetic braking would provide the best constraint to date on the asymptotic curvature radius of the extra dimension of the order of $35~{\rm \mu m}$.
   \end{enumerate}

\begin{acknowledgements}
I would like to thank Dimitrios Psaltis for carefully reading the manuscript. This work was supported by the NSF CAREER award NSF 0746549.
\end{acknowledgements}

\end{document}